\RequirePackage{fix-cm}
\documentclass[twocolumn,epjc3,a4paper]{svjour3-preprint}
\smartqed  
\RequirePackage{graphicx}
\RequirePackage{flushend}
\RequirePackage{url,cite,xspace,amsmath,amssymb}
\RequirePackage[colorlinks,citecolor=blue,urlcolor=blue,linkcolor=blue]{hyperref}
\RequirePackage[mathlines]{lineno}
\newcommand{\none}{\ensuremath{\tilde{\chi}_1^0}\xspace}
\newcommand{\ntwo}{\ensuremath{\tilde{\chi}_2^0}\xspace}
\newcommand{\chone}{\ensuremath{\tilde{\chi}_1^{\pm}}\xspace}

\newcommand{\chonep}{\ensuremath{\tilde{\chi}_1^+}\xspace}
\newcommand{\chonem}{\ensuremath{\tilde{\chi}_1^-}\xspace}
\newcommand{\glu}{\ensuremath{\tilde{g}}\xspace}
\newcommand{\sq}{\ensuremath{\tilde{q}}\xspace}

\newcommand{\stauu}{\ensuremath{\tilde{\tau}}\xspace}
\newcommand{\stau}{\ensuremath{\tilde{\tau}_1}\xspace}
\newcommand{\staur}{\ensuremath{\tilde{\tau}_{\mathrm R}}\xspace}
\newcommand{\gra}{\ensuremath{\tilde{G}}\xspace}
\newcommand{\piplus}{\ensuremath{\pi^{\pm}}\xspace}

\newcommand{\ifb}{\mbox{fb$^{-1}$}\xspace}
\newcommand{\mev}{\mbox{MeV}\xspace}
\newcommand{\gev}{\mbox{GeV}\xspace}
\newcommand{\tev}{\mbox{TeV}\xspace}
\newcommand{\pev}{\mbox{PeV}\xspace}
\newcommand{\eev}{\mbox{EeV}\xspace}
\newcommand{\met}{\ensuremath{E_{\mathrm T}^\text{miss}}\xspace}

\newcommand{\madnlo}{\textsc{MadGraph5}\_a\textsc{MC@NLO}\xspace}
\newcommand{\pyt}{\textsc{Pythia}\xspace}

\DeclareGraphicsExtensions{.pdf,.png,.jpg,.jpeg}
\graphicspath{{./}{./figs/}}

\journalname{Eur. Phys. J. C}
\begin{document}
\preprint{IFIC/19-58 \\ FTUV-19-1216\\ KCL-PH-TH/2019-93}
\title{Prospects for discovering supersymmetric long-lived particles with MoEDAL}

\author{D.~Felea\thanksref{e1,bucharest}
\and J.~Mamuzic\thanksref{e2,valencia}
\and R.~Mase\l{}ek\thanksref{e3,warsaw}
\and N.~E.~Mavromatos\thanksref{e4,london}
\and V.~A.~Mitsou\thanksref{e5,valencia}
\and J.~L.~Pinfold\thanksref{e6,alberta}
\and R.~Ruiz~de~Austri\thanksref{e7,valencia}
\and K.~Sakurai\thanksref{e8,warsaw}
\and A.~Santra\thanksref{e9,valencia}
\and O.~Vives\thanksref{e10,valencia,valencia2}
}

\thankstext{e1}{e-mail: daniel.felea@cern.ch}
\thankstext{e2}{e-mail: judita.mamuzic@ific.uv.es}
\thankstext{e3}{e-mail: r.maselek@uw.edu.pl}
\thankstext{e4}{e-mail: nikolaos.mavromatos@kcl.ac.uk}
\thankstext{e5}{e-mail: vasiliki.mitsou@ific.uv.es}
\thankstext{e6}{e-mail: jpinfold@ualberta.ca}
\thankstext{e7}{e-mail: rruiz@ific.uv.es}
\thankstext{e8}{e-mail: kazuki.sakurai@fuw.edu.pl}
\thankstext{e9}{e-mail: arka.santra@cern.ch}
\thankstext{e10}{e-mail: oscar.vives@uv.es}


\institute{Institute of Space Science, P.O.\ Box MG~23, 077125, Bucharest -- M\u{a}gurele, Romania\label{bucharest} 
\and Instituto de F\'isica Corpuscular (IFIC), CSIC -- Universitat de Val\`encia, C/ Catedr\'atico Jos\'e Beltr\'an 2, E-46980 Paterna (Valencia), Spain\label{valencia} 
\and Institute of Theoretical Physics, Faculty of Physics, University of Warsaw, ul.\ Pasteura 5, PL02093 Warsaw, Poland\label{warsaw}  
\and Theoretical Particle Physics and Cosmology Group, Department of Physics, King's College London, Strand, London WC2R~2LS, UK\label{london}  
\and Physics Department, University of Alberta, Edmonton Alberta T6G~2E4, Canada\label{alberta}  
\and Departament de F\'isica Te\`orica, Universitat de Val\`encia, C/ Dr.\ Moliner 50, E-46100 Burjassot (Valencia), Spain\label{valencia2} 
}
\date{Received: date / Accepted: date}
%
\maketitle

\begin{abstract}
We present a study on the possibility of searching for long-lived supersymmetric partners with the MoEDAL experiment at the LHC. MoEDAL is sensitive to highly ionising objects such as magnetic monopoles or massive (meta)stable electrically charged particles. We focus on prospects of directly detecting long-lived sleptons in a phenomenologically realistic model which involves an intermediate neutral long-lived particle in the decay chain. This scenario is not yet excluded by the current data from ATLAS or CMS, and is compatible with astrophysical constraints. Using Monte Carlo simulation, we compare the sensitivities of MoEDAL versus ATLAS in scenarios where MoEDAL could provide discovery reach complementary to ATLAS and CMS, thanks to looser selection criteria combined with the virtual absence of background. It is also interesting to point out that, in such scenarios, in which charged staus are the main long-lived candidates, the relevant mass range for MoEDAL is compatible with a potential role of Supersymmetry in providing an explanation for the anomalous events observed by the ANITA detector.
\end{abstract}

\section{Introduction}\label{sc:intro}

Supersymmetry (SUSY)~\cite{Wess:1974tw,Salam:1974ig,Ferrara:1974pu,Martin:1997ns}, and its local ``gauged'' version, through its embedding in supergravity scenarios (SUGRA), is a well-motivated extension of the Standard Model (SM) from a theoretical point of view, which assigns to each SM field a superpartner field with a spin differing by a half unit. SUSY provides elegant solutions to several open issues in the SM, such as the hierarchy problem, the identity of dark matter, and grand unification. Its nondiscovery, as yet, at current colliders sets strong constraints to minimal versions, such as the minimal supersymmetric standard model (MSSM) and its minimal $N=1$ supergravity extensions (mSUGRA). There are compelling arguments that SUSY might still be discovered in the foreseeable future~\cite{Ellis:2019ujs}, in the sense that there are still unexplored regions in the available parameter space of current collider searches. The latter can be probed either by testing non-conventional models, for instance, $R$-parity violating (RPV) models~\cite{Barger:1989rk,Dreiner:1997uz,Barbier:2004ez}, which incidentally may provide elegant explanations for the origin of neutrino masses~\cite{Hirsch:2000ef}, or through signatures that have not been previously explored in depth, such as those due to the existence of long-lived particles (LLPs), which are predicted in some SUSY scenarios~\cite{Fairbairn:2006gg}. The LLPs may either decay within the typical volume of an LHC detector or may be sufficiently long-lived ((meta)stable) so as to traverse the entire detector without decaying. In the former case, it may give rise to displaced vertices~\cite{Aaboud:2018aqj,Sirunyan:2018vlw} or disappearing tracks~\cite{Aaboud:2017mpt,Sirunyan:2018ldc}. Here we focus on ``collider-stable'' particles and more precisely on heavy, stable charged particles (HSCPs),\footnote{If the stable particle is neutral, hence only weakly interacting, such as the \none, its signature of large missing transverse momentum is typical for SUSY searches and therefore it is not discussed in the context of LLPs.} predicted in some SUSY models to be specified below, that may give rise to anomalous ionisation detectable by the MoEDAL detector. 

HSCPs may be observed in detectors optimised for signals of high ionisation, both in collider experiments~\cite{Lee:2018pag,Alimena:2019zri} as well as in cosmic observatories~\cite{Burdin:2014xma}. The general-purpose ATLAS and CMS experiments at the Large Hadron Collider (LHC), in particular, have searched for and have constrained theoretical scenarios that predict highly ionising particles (HIPs) already since Run~1~\cite{Khachatryan:2011ts,Aad:2011yf}. Besides them, dedicated detectors are being proposed to explore these less-constrained manifestations of physics beyond the SM~\cite{Alimena:2019zri}. Among them, the Monopole and Exotics Detector At the LHC (MoEDAL)~\cite{Pinfold:2009oia} is the only one in operation as of today. It is specially designed to detect HIPs such as magnetic monopoles and HSCPs, covering a wide spectrum of theoretical models~\cite{Acharya:2014nyr}, in a manner complementary to CMS and ATLAS~\cite{DeRoeck:2011aa}. 

It is the purpose of this article to discuss the supersymmetry discovery potential of MoEDAL by presenting a SUSY model case study which clearly demonstrates the complementarity of this detector to that of ATLAS and CMS searches. We study a specific supersymmetric model predicting HSCPs and determine the relevant parameter range in terms of masses and lifetimes for which the MoEDAL detector could observe a possible signal. 

As an interesting byproduct of our analysis, we also present a brief discussion  on the anomalous air shower events observed by the ANITA Collaboration~\cite{Gorham:2016zah,Gorham:2018ydl},
putting emphasis on the fact that the range of HSCP parameters probed by MoEDAL can be in the interesting regime of providing explanations for those events based on supersymmetric models~\cite{Albuquerque:2003mi,Ando:2007ds,Dudas:2018npp,Fox:2018syq,Connolly:2018ewv,Collins:2018jpg,Anchordoqui:2019utb,Cline:2019snp,Huang:2018als}. Astrophysical explanations of these events are in tension with IceCube data~\cite{Aartsen:2017mau,Aartsen:2020vir}, strengthening the possibility for an origin from physics Beyond the SM (BSM). However, we stress that this connection is only mentioned here as a potentially interesting additional motivation for our analysis. Although elegant, by no means we wish to promote the supersymmetric origin of these events here,  since more mundane explanations are possible~\cite{Shoemaker:2019xlt}. 

The structure of this paper is as follows. In Section~\ref{sc:hscp}, we discuss SUSY models predicting HSCPs, also reviewing their current experimental constraints set from LHC experiments. An overview of the MoEDAL detector components and analysis techniques, emphasising the complementarity to the  approach followed in ATLAS and CMS is given in section~\ref{sc:moedal}. In Section~\ref{sc:direct}, we study the SUSY HSCP direct production kinematics relevant to MoEDAL. Section~\ref{sc:realistic} presents results from a case study of a simplified topology where MoEDAL can be sensitive to regions of the parameter space different than the respective of ATLAS and CMS. In Section~\ref{sc:anita}, we connect our results in this article with potential supersymmetry-inspired explanations of the ANITA anomalous events. We finally close the report with some concluding remarks and an outlook in Section~\ref{sc:concl}.

\section{HSCPs and SUSY at the LHC}\label{sc:hscp}

In supersymmetric models, various instances of sparticles may emerge as HSCPs. Considering its detectors position in the cavern and its sensitivity to slow-moving particles, MoEDAL may detect HSCPs with proper lifetimes  $c\tau\gtrsim1$~m. 

\noindent {\bf Sleptons.} They may be long-lived as next-to-the-lightest SUSY partners (NLSPs) decaying to a gravitino (\gra) or a neutralino (\none) LSP. In gauge-mediated supersymmetry breaking (GMSB) scenarios, the \stau NLSP decays to \gra may be suppressed due to the ``weak'' gravitational interaction~\cite{Hamaguchi:2006vu}, remaining partially compatible with constraints on the dark-matter abundance in super-weakly interacting massive particle scenarios~\cite{Feng:2015wqa}. In other cases, such as the  co-annihilation region in the constrained MSSM, the most natural candidate for the NLSP is the lighter \stau, which could be long lived if the mass splitting between the \stau and the \none is smaller than the $\tau$-lepton mass~\cite{Jittoh:2005pq,Kaneko:2008re,Feng:2015wqa}.\footnote{For light-flavour sleptons, the condition $m_{\tilde \ell} - m_{\none} < m_{\ell}$ requires much higher degree of fine tuning and long-lived light-flavour sleptons are not usually considered. In addition, the \stau is typically lighter than other sleptons due to the effect of the $\tau$ Yukawa coupling in renormalisation group equations evolution, hence it is likely more accessible in collider experiments.}
This region is one of the most favoured by the measured dark-matter relic density~\cite{Ellis:2003cw}.

\noindent {\bf R-hadrons.} They are formed by hadronised metastable gluinos, light-flavour squarks, stops or sbottoms. Gluino R-hadrons arise in Split SUSY~\cite{ArkaniHamed:2004fb,Giudice:2004tc} due to the extremely heavy squarks that suppress strongly \glu decays to \sq and quarks~\cite{ArkaniHamed:2004fb,ArkaniHamed:2004yi}. Other models, such $R$-parity-violating SUSY~\cite{Evans:2012bf} or gravitino dark matter~\cite{DiazCruz:2007fc}, could produce a long-lived squark that would also form an R-hadron. 

\noindent {\bf Charginos.} They may be very long-lived as lightest supersymmetric particles (LSPs) in RPV models with relatively weak RPV couplings~\cite{Bomark:2014rra} or as NLSPs in gravitino LSP scenarios~\cite{Kribs:2008hq}, thus making their detection possible due to high ionisation. Long lifetime may also be due to mass degeneracy with the \none LSP, e.g., in anomaly-mediated symmetry breaking (AMSB) scenarios~\cite{Giudice:1998xp,Randall:1998uk} or in the focus-point region of the mSUGRA parameter space~\cite{Gladyshev:2008ag}. However, in the latter cases the \chone lifetime is moderately long, leading to decays within the detectors to a soft \piplus and a \none, which are constrained by searches for disappearing tracks~\cite{Aaboud:2017mpt,Sirunyan:2018ldc}. 
ATLAS and CMS have searched for stable sleptons, R-hadrons and charginos using anomalously high energy deposits in the silicon tracker and timing measurements in the calorimeters and the muon system. The most recent ATLAS analysis~\cite{Aaboud:2019trc} has set the most stringent limits with 36.1~\ifb of $pp$ collisions at 13~\tev, while CMS has used 2.5~\ifb so far~\cite{Khachatryan:2016sfv}.  The ATLAS bounds at 95\% confidence limit (CL) are 2000~\gev for gluino R-hadrons, 1250~\gev for sbottom R-hadrons, 1340~\gev for stop R-hadrons,  430~\gev for sleptons and 1090~\gev for charginos with sufficiently long lifetime. In refs.~\cite{ATLASsummary,CMSsummary}, constantly updated summary plots of ATLAS and CMS analyses results pertaining to HSCPs are provided. For comprehensive and recent reviews on LHC past, current and future LLP searches, the reader is referred to Refs.~\cite{Lee:2018pag,Alimena:2019zri}. 

\section{MoEDAL complementarity to ATLAS and CMS}\label{sc:moedal}

The MoEDAL experiment~\cite{Pinfold:2009oia} is installed around the intersection region at LHC Point~8 (IP8) in the LHCb vertex locator cavern. It is a unique and largely passive detector comprising different detector technologies, highlighted below.

The MoEDAL main subdetectors are made of a large array of CR-39, Makrofol\textsuperscript{\small\textregistered} and Lexan{\small\sf\texttrademark} nuclear track detector (NTD) panels surrounding the intersection area. The passage of a HIP through the plastic sheet is marked by an invisible damage zone along the trajectory, which is revealed as a cone-shaped etch-pit when the plastic detector is chemically etched. Then the detector is scanned in search of aligned etch pits in multiple sheets. The NTDs of MoEDAL have a threshold of $z/\beta\sim5$, where $z$ is the charge and $\beta=v/c$ the velocity of the incident particle.

A unique feature of the MoEDAL detector is the use of magnetic-monopole trappers (MMTs) to capture charged HIPs. In the case of monopoles, the high magnetic charge implies a strong magnetic dipole moment, which may result in strong binding of the monopole with the nuclei of the aluminium MMTs. In such a case, the presence of a trapped monopole would be detected in a superconducting magnetometer through the induction technique~\cite{Acharya:2019vtb}. 

In addition, the MMTs may also capture HSCPs, which can only be observed through the detection of their decaying products. To this effect, the MoEDAL Collaboration is planning the MoEDAL Apparatus for detecting extremely Long Lived particles (MALL)~\cite{Alimena:2019zri}. In this case, MoEDAL MMTs, after they have been scanned through a magnetometer to identify any trapped monopole, will be installed underground to be monitored for the decay of captured particles. MALL is expected to be sensitive to charged particles and to photons, with energy as small as 1~\gev. 

Another handle on constraining SUSY LLPs can be provided by the MoEDAL Apparatus for Penetrating Particles (MAPP), which is designed to search for milli-charged particles of electric charge $\gtrsim 0.001e$,  and for new long-lived neutrals decaying to charged SM particles~\cite{Pinfold:2019zwp}. This subdetector is going to be fully operational during the LHC Run~3, along with the baseline MoEDAL detectors. It will be installed $\sim 30~{\mathrm m}$ from the interaction point, thus it will be sensitive to very delayed decays of neutral particles such as neutralinos in RPV scenarios~\cite{Dercks:2018eua,Dercks:2018wum}.

Given the unique design of the MoEDAL subsystems, the complementary aspects of MoEDAL to ATLAS and CMS, as far as HSCPs are concerned, come as no surprise. MoEDAL is practically ``time-agnostic'' due to the passive nature of its detectors. Therefore, signal from very slowly moving particles will not be lost due to arriving in several consecutive bunch crossings. Moreover, ATLAS and CMS carry out trigger-based searches for LLPs, which may trigger on accompanying ``objects'', such as missing transverse momentum, \met  (see, e.g., Refs.~\cite{Aaboud:2019trc,Khachatryan:2016sfv}). Alternatively, specialised triggers have been developed and applied, which have usually relatively low efficiency. For instance, the recent magnetic monopole ATLAS search~\cite{Aad:2019pfm} utilises a trigger based on the tracker high-threshold hit capability with a level-1 trigger efficiency ranging from 10\% -- 60\%. In another example, a late-muon trigger aiming at recovering efficiency for slow particles by considering two consecutive bunch crossings, which was partly active in ATLAS Run~2, is expected to have an efficiency of $\lesssim 15\%$ for \glu R-hadrons~\cite{Heinrich:2018pkj}. For comparison, we note here that the triggers used in SUSY searches involving promptly decaying sparticles, have typically very high efficiency, as e.g.\ in Refs.~\cite{Sirunyan:2019xwh,Aaboud:2017vwy,ATLAS:2019vcq,CMS:2019tlp,Sirunyan:2019iwo,Aad:2019qnd}, where the \met, single-lepton, photon triggers used are more than 95\% efficient. 

MoEDAL, on the other hand, is primarily limited by the lower luminosity delivered at IP8, by the geometrical acceptance of the detectors, especially the MMTs, and by the requirement of passing the $z/\beta$ threshold of NTDs. In general, ATLAS and CMS have demonstrated their ability to cover high velocities, while MoEDAL is sensitive to lower ones $\beta \lesssim 0.2$. Typically $\beta \gtrsim 0.5$ is a safe limit for ATLAS and CMS, due to hit/track information passing to a different bunch crossing, thus making it very difficult to reconstruct, if at all possible.

Both ATLAS and CMS have to select the interesting events out of a large background of known SM processes which may fake signal events. To suppress this background, they have to apply offline cuts that unavoidably limit the efficiency of LLP detection, hence reducing the parameter space probed by ATLAS and CMS. On the other hand, MoEDAL has practically no background and requires no trigger or selection cuts to detect a HIP, therefore it may detect particles that may escape detection at other LHC experiments.

Regarding particles stopped in material and their subsequent decays, different approaches are followed. ATLAS and CMS look in empty bunch crossings for decays of trapped particles into jets~\cite{Aad:2013gva,Sirunyan:2017sbs}, with background coming from beam-halo events and cosmic muons. MALL, on the other hand, is currently planned to be installed in one of the underground galleries of IP8 and its background is expected to come mainly from cosmic rays. The probed lifetimes should be larger than those constrained by ATLAS/CMS --- up to $\sim\!10$~years according to initial estimates --- due to the unlimited monitoring time. 

\section{Direct production of metastable sparticles at the LHC}\label{sc:direct}

In this study, we discuss the kinematics of metastable sparticles in 13~\tev $pp$ collisions, focusing on their velocity $\beta$, which is the figure of merit for MoEDAL. Throughout our study, we use \madnlo~\cite{Alwall:2014hca} and \pyt~8~\cite{Sjostrand:2007gs} for Monte Carlo simulation. The $\beta$ distributions in the direct \staur pair production are shown in Fig.~\ref{fg:stau-direct} for various \staur masses. The fraction of events with $\beta \lesssim 0.2$, i.e.\ within the range of NTD sensitivity, only becomes significant for large \staur masses of ${\mathcal O}(1~\tev)$. In this mass range, the cross section is very low, as shown in Fig.~\ref{fg:xs}, making the possibility for \staur detection in the NTDs marginal.
 
\begin{figure}[htb]
\centering
\includegraphics[width=0.95\linewidth]{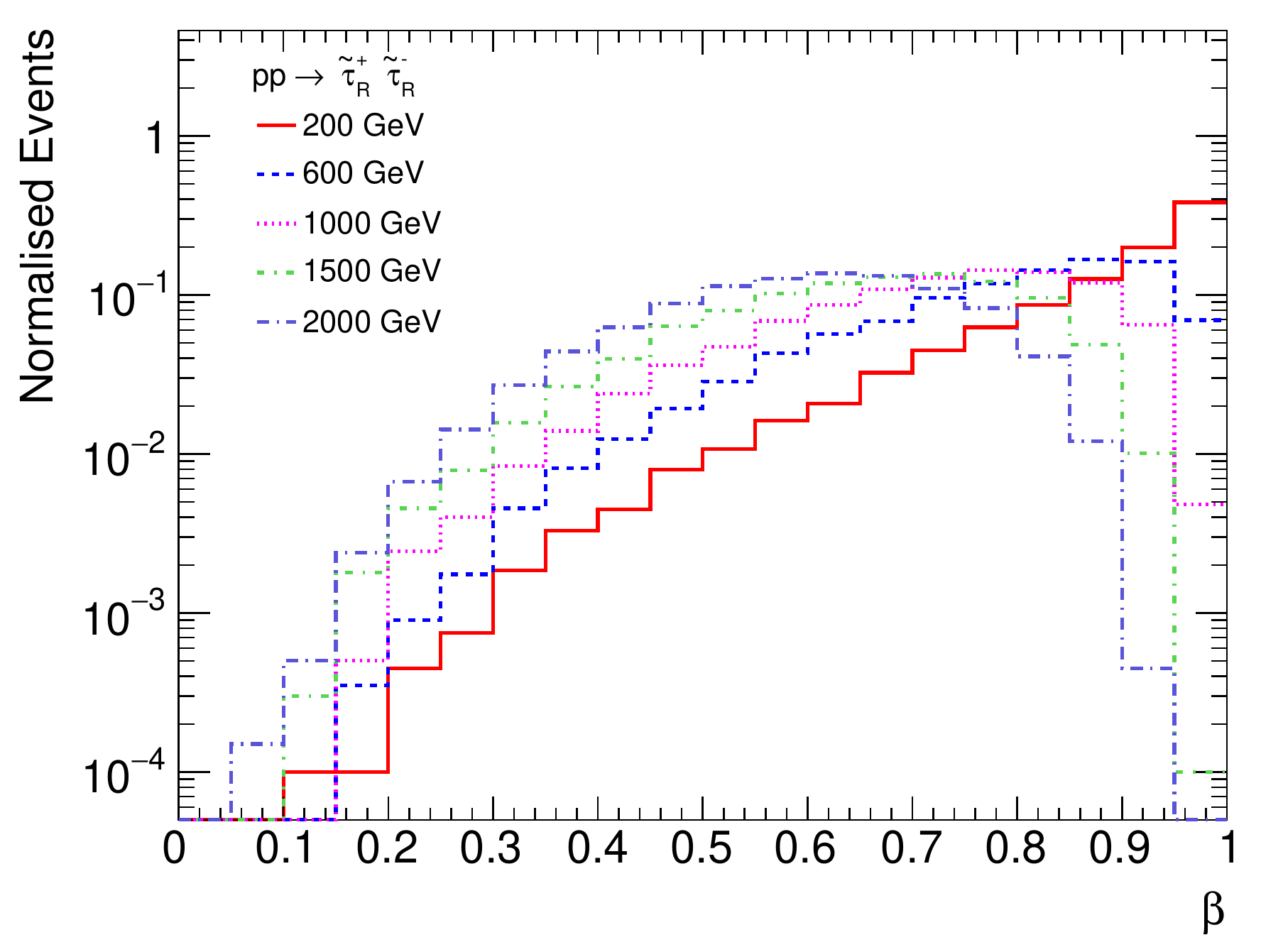}
\caption{\label{fg:stau-direct} Stau velocity distributions for $\staur^+\staur^-$ direct production in 13~\tev $pp$ collisions for \staur masses between 200~\gev and 2~\tev.}
\end{figure}

\begin{figure}[htb]
\centering
\includegraphics[width=0.95\linewidth]{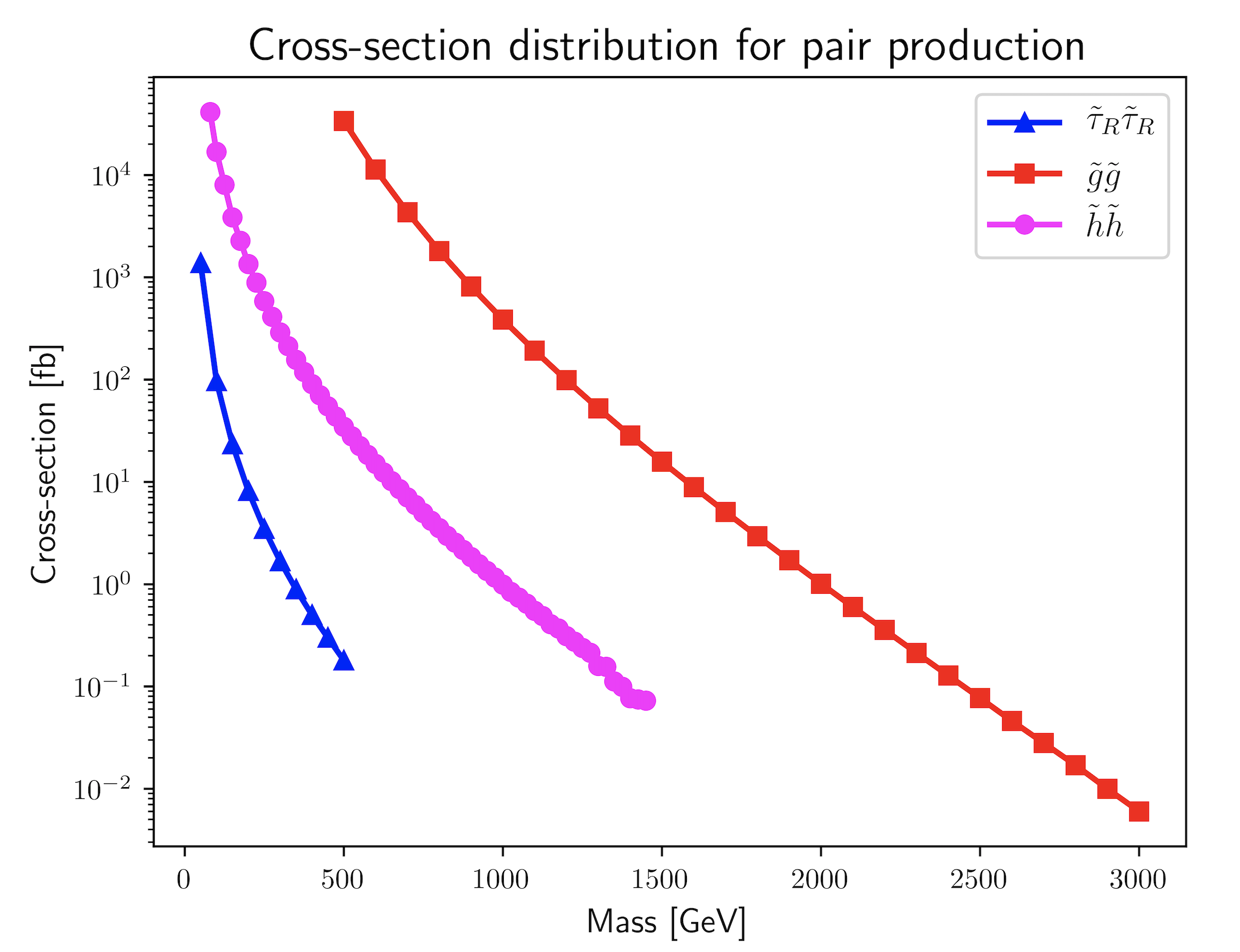}
\caption{\label{fg:xs} The cross sections for pair production at the 13~\tev LHC of staus (blue) and higgsinos (magenta) at NLO+NLL level and for gluinos (red) at NNLO\textsubscript{approx}+NNLL precision taken from Ref.~\cite{sxwg}.}
\end{figure}

We have also simulated the direct pair production of higgsinos ($\none\,\chone$, $\ntwo\,\chone$) and gluinos ($\glu\,\!\glu$), besides that of staus ($\staur^+\staur^-$). As evident from their $\beta$ distributions in Fig.~\ref{fg:fermion-boson}, fermions (gluinos, hisggsinos) are slower than bosons (staus) and, therefore, have larger ionisation energy loss. This is because the dominant channel is an $s$-channel spin-1 gauge boson $(Z^*/\gamma^*)$ exchange with $q\bar{q}$ initial states. The gauge bosons are transversely polarised due to helicity conservation in the initial vertex, so the final state must have a total non-zero angular momentum. The scalar (spinless) pair production (\stauu) undergoes a $p$-wave suppression, i.e.\ the production cross section vanishes as the \stauu velocity goes to zero to conserve angular momentum. No such suppression exists in the fermion (spinful) case. 

\begin{figure}[htb]
\centering
\includegraphics[width=0.95\linewidth]{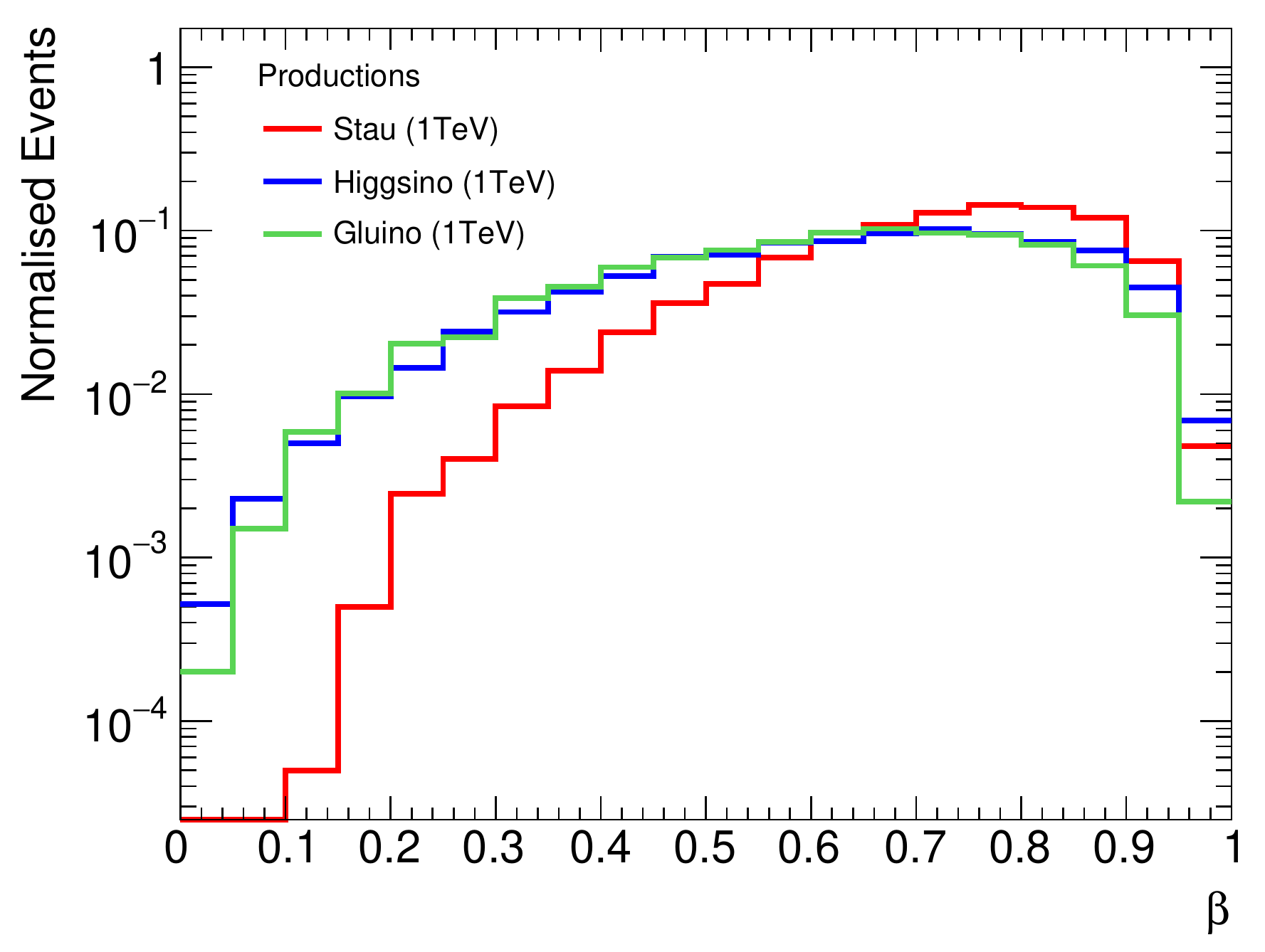}
\caption{\label{fg:fermion-boson} Comparison of velocity distributions between staus, higgsinos and gluinos of the same mass ($1~\tev$) produced directly in pairs in 13~\tev $pp$ collisions.}
\end{figure}

For comparison, we show the cross sections for stau, higgsino and gluino pair production at the 13~\tev LHC in Fig.~\ref{fg:xs} with values  obtained from Ref.~\cite{sxwg}. The higgsino case includes all production modes, $\none\ntwo + \none\chone + \ntwo\chone + \chonep\chonem$, where these gauginos are assumed to be mass degenerate.\footnote{Here we consider the case where the charged component of the higgsino $\chone$ is slightly lighter than the neutral components, $\none$ and $\ntwo$, such that the neutral states decay into the stable $\chone$ before reaching and detected by MoEDAL's NTDs. This small mass splitting is ignored in the total cross-section shown in Fig.~\ref{fg:xs}.} The stau and higgsino cross sections are calculated at next-to-leading order (NLO) plus next-to-leading logarithmic (NLL) precision, while for gluinos the precision is at the approximate next to NLO (NNLO\textsubscript{approx}) plus next to NLL (NNLL). Between higgsinos and gluinos, the latter would be preferable in this context as they are typically produced more abundantly. 

To conclude, gluino pair direct production should serve as the best scenario for MoEDAL, since they are heavy fermions with large cross section. In the following, we discuss the lightest \stau as a HSCP produced in \glu cascade decays, leaving the study of \glu R-hadrons for the future.\footnote{In the following section, we do not assume that the lightest stau is dominantly a right-handed partner. This is because, unlike the direct stau pair production studied in this section, the final-state staus are produced via decay of gluinos and the signal yield is independent of the left-right mixing in the stau sector. In Section~\ref{sc:anita}, we again assume the lightest stau is dominantly a  right-handed one, when discussing the ANITA events in the context of a gauge mediated SUSY breaking scenario. } 

\section{MoEDAL sensitivity to staus}\label{sc:realistic}

Some preliminary studies on MoEDAL reach in comparison with CMS projections showed that MoEDAL can be complementary to ATLAS/CMS despite the lower luminosity available at IP8~\cite{Sakurai:2019bac}.\footnote{Indicatively, in Run~2 the delivered luminosity at IP8 (LHCb and MoEDAL) was a factor of $\sim20$ times less than that recorded at IP1 (ATLAS) and IP5 (CMS). The LHCb full software trigger, part of the Phase-1a Upgrade for Run~3~\cite{LHCbCollaboration:2014vzo}, will allow an increased collision rate at IP8 leading to an expected integrated luminosity of 30~\ifb for Run~3, compared to 300~\ifb for ATLAS/CMS, thus reducing this factor to $\sim 10$. } That study was using a simplistic description of the MoEDAL NTDs and the CMS efficiencies for HSPCs published in Ref.~\cite{Khachatryan:2015lla}, extracted to re-interpret a previous HSCP search performed by CMS~\cite{Chatrchyan:2013oca} in specific supersymmetric models at energies of 7 and 8~\tev. 

As discussed earlier, we concentrate our efforts on heavy long-lived sparticles with a large production cross section that in addition respect present bounds. Therefore, we do not only study the MoEDAL sensitivity, but we also compare it with the latest HSCP search conducted by ATLAS~\cite{Aaboud:2019trc}. As can be seen in Fig.~\ref{fg:fermion-boson}, the fraction of events with $\beta \lesssim 0.2$, i.e. within the NTD sensitivity, is only $\sim 1\%$ even for gluinos. Because of this and due to the lower luminosity delivered to MoEDAL, ATLAS and CMS in general provide much better sensitivities for HSCPs. We therefore focus on a particular scenario where ATLAS and CMS may loose their sensitivity while MoEDAL retains it.

\subsection{Model description}\label{sc:model}

In the ATLAS and CMS HSCP searches, multiple hits in the (innermost) pixel detector are required to ensure good track reconstruction of \emph {charged\ } particles. However, the presence of a \emph{neutral\ } long-lived sparticle in the cascade decay may dissatisfy this selection criterion, thus limiting the acceptance of such model. This is expected to become evident in particular in regions of the parameter space with large lifetime of this intermediate particle.

This observation leads us to consider a gluino pair production ($pp \to \tilde g \tilde g$) followed by the prompt decay of gluino into a long-lived neutralino plus two quark jets; $\tilde g \to \tilde \chi_1^0 q \bar q$. We assume that the long-lived neutralino may decay, after travelling $\sim 1$\,m, into an off-shell tau-lepton plus a metastable stau, $\tilde \chi_1^0 \to \tilde \tau_1 \tau^*$, due to a very small mass splitting: $\delta m = m_{\tilde \chi_1^0} - m_{\tilde \tau_1} \lesssim m_\tau$. 
\begin{linenomath*}
\begin{align}\label{eq:cascade}
 p p \to \glu \glu  &\to \left ( \none jj\right) \left ( \none jj\right) \nonumber \\ &\to \left  ( \stauu_{\text{1,dv}} \tau^*_{\text{dv}} jj\right) \left ( \stauu_{\text{1,dv}} \tau^*_{\text{dv}} jj\right).
\end{align}
\end{linenomath*} 
The subscript ``$\text{dv}$'' indicates that the particles originate from a displaced vertex. The \none lifetime depends on its mass difference with the \stau, as $\propto (\delta m)^6$ in 3-body decays~\cite{Kaneko:2008re,Khoze:2017ixx}. So, the lifetime can be tuned from $\sim 10^{-9}$~s for $\delta m\sim 1.7~\gev$ to $\sim 10^6$~s  for $\delta m\sim 500~\mev$, which would imply decay lengths from 10~cm to 100~m.

Finally, the metastable staus may decay, after passing through the detector, into $\tau$'s and other SM particles via very small RPV couplings, when present with a $\stauu$ LSP, or into a $\tau$ and \gra LSP, via gravitational interaction if they are the NLSPs. All other supersymmetric particles are decoupled and they do not play a role in the following analysis. 

\subsection{ATLAS analysis recasting and other constraints}\label{sc:recast}

The latest HSCP search by CMS~\cite{Khachatryan:2016sfv} uses only 2.5~\ifb of $pp$ collision data at 13~\tev. Since the analysis design and selection cuts are very similar to those of ATLAS, we only focus on Ref.~\cite{Aaboud:2019trc} by ATLAS, which has analysed more data: 36.1~\ifb from LHC Run~2. However, the CMS results should also be relevant for the same dataset size.

In the cascade decay~\eqref{eq:cascade}, with a long \none lifetime ($c \tau_{\none} \sim 1$~m), multiple pixel hits cannot be expected because what is travelling in the pixel detector is the invisible neutralino. The ATLAS analysis, in particular, requires seven pixel hits. The probability (per particle) of having all pixel hits for our simplified model is proportional to the probability of the \none decaying before reaching the pixel detector, that is
\begin{linenomath*}
\begin{equation}
P_{\text{pixel}} = 1 - \exp \left(-\frac{L_{\text{pixel}}}{ \beta \gamma c \tau_{\none} \sin\theta} \right),
\end{equation}
\end{linenomath*}
where $\gamma \equiv \frac{1}{\sqrt{1 - \beta^2}}$ with $\beta$ being the \none velocity, $\theta$ ($\theta \in [0, \pi/2]$) is the angle between the \none momentum and the beam axis, $L_{\text{pixel}}/\sin\theta$ is the distance between the interaction point to the pixel detector and $L_{\text{pixel}} = 50.5$\,mm is the minimum distance between the interaction point and the first layer of the pixel detector (at $\theta = \pi/2$). We see that $P_{\text{pixel}} \ll 1$ for $c \tau_{\none} \gg L_{\text{pixel}}$. 

In recasting the latest ATLAS HSCP search, we closely follow the recipe provided in the HEPData record~\cite{hepdata} of Ref.~\cite{Aaboud:2019trc}, where various information, such as the trigger efficiency and the efficiency maps for signal reconstruction, are also given. We estimated the current limit in terms of $m_{\tilde g}$ and $c \tau_{\tilde \chi_1^0}$ by multiplying $P_{\text{pixel}}$ with the signal efficiency obtained by the official recasting procedure.

Other analyses that may potentially constrain the model under study are the ones targeting displaced jets (also sensitive to hadronic $\tau$'s)~\cite{Aaboud:2017iio,Aaboud:2018aqj,Sirunyan:2018vlw} or displaced leptons (from leptonic $\tau$ decays)~\cite{Aaboud:2018jbr,CMS:2014hka}. Due to the current unavailability of recasting instructions and related tools for these analyses --- which is due to the unconventional detector utilisation --- we do not consider them here.

\subsection{MoEDAL detector geometry and response}\label{sc:detector}

We estimate the MoEDAL detection sensitivity of this gluino cascade scenario as accurately as possible without using the detailed full \textsc{Geant4} simulation for the detector response. In this study at a first stage, we consider the \mbox{Run-2} (2015-2018) NTD deployment shown in Fig.~\ref{fg:geom}. The geometrical acceptance, i.e.\ the fraction of the solid angle covered by the NTD panels, of this configuration is $\sim 20$\%. In order for the staus in the cascade chain to be detected by MoEDAL, the neutralino must decay and produce a stau before reaching a NTD panel, and the produced stau must hit the NTD panel. Since the mass splitting between $\tilde \chi_1^0$ and $\tilde \tau_1$ is assumed to be much less than $m_{\tau} = 1.777~\gev$, the $\tilde \tau_1$ and $\tilde \chi_1^0$ are travelling almost in the same direction. For a given neutralino momentum, ${\bf p}_{\tilde \chi_1^0}$, the probability for the stau to hit a NTD panel is given by
\begin{linenomath*}
\begin{equation}
P_{\rm NTD}({\bf p}_{\tilde \chi_1^0}) = \omega({\bf p}_{\tilde \chi_1^0}) \left[ 1 - \exp\left( \frac{L_{\rm NTD}({\bf p}_{\tilde \chi_1^0})}{\beta \gamma c \tau_{\tilde \chi_1^0}} \right) \right],
\end{equation}
\end{linenomath*}
where $\omega({\bf p}_{\tilde \chi_1^0}) = 1$ if there is a NTD panel in the direction of ${\bf p}_{\tilde \chi_1^0}$ and $0$ otherwise and $L_{\rm NTD}({\bf p}_{\tilde \chi_1^0})$ is the distance to the NTD panel in the direction of ${\bf p}_{\tilde \chi_1^0}$. On average $L_{\rm NTD} \sim 2$\,m. 

\begin{figure}[htb]
\centering
\includegraphics[width=0.95\linewidth]{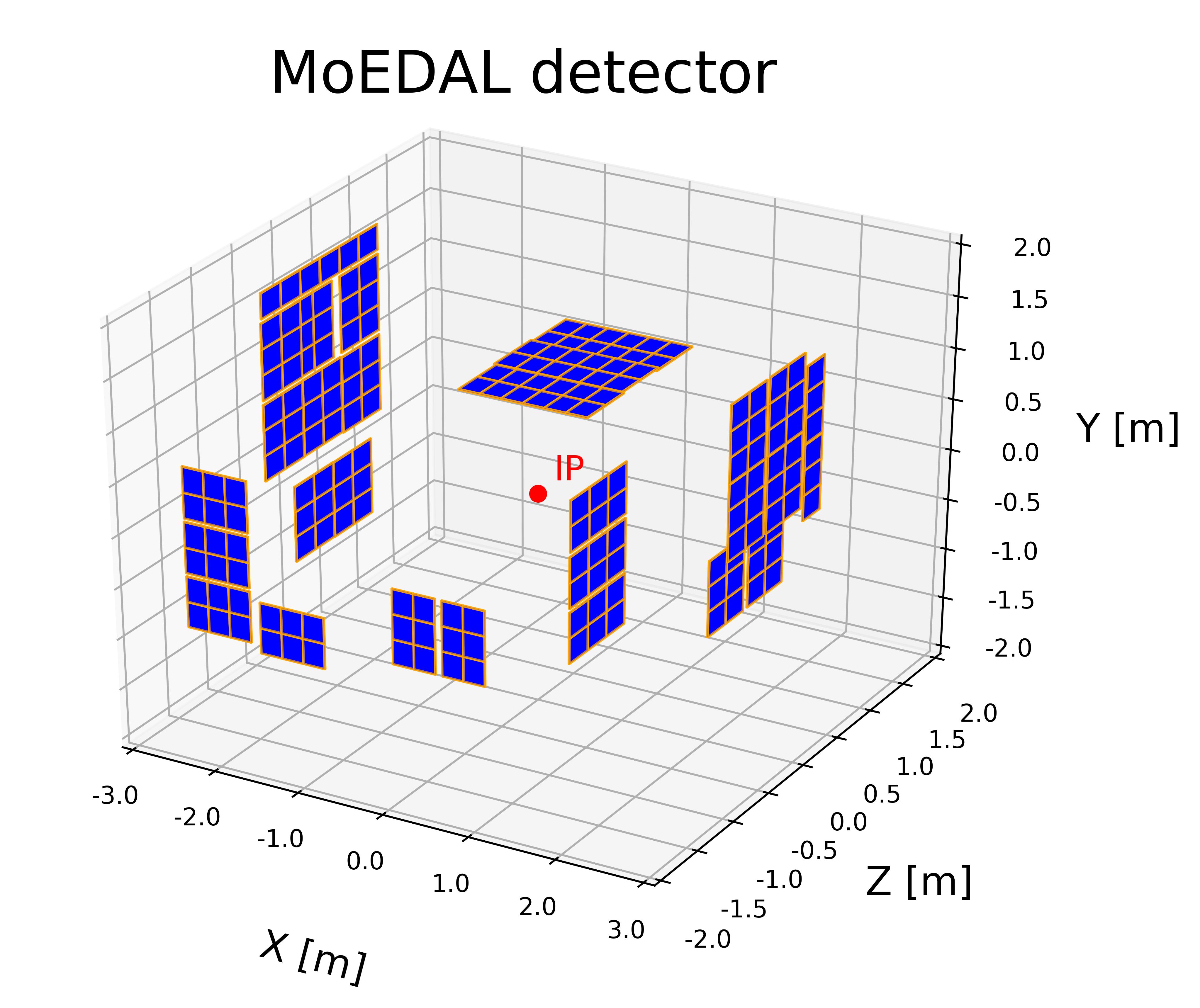}
\caption{\label{fg:geom} The Run-2 NTD deployment of MoEDAL. NTD modules are depicted as thin blue plates with orange edges. The red point at the centre represents the interaction point. The $z$-axis is along the beams and the $y$-axis indicates the vertical direction.}
\end{figure}

When the stau hits the NTD panel, its detectability depends on the incidence angle between the stau and the NTD panel as well as the stau's velocity. This is because if the incidence is shallow and the velocity is large, the etch-pit is tilted and small~\cite{Patrizii:2010jla,Manzoor:2007}. Such an etch-pit will not survive when the surface of NTD panel is chemically etched and removed. For any given $\beta$, the stau is detected only when its incidence angle to the NTD panel, $\delta$ ($\delta \in [0^{\circ}, 90^{\circ}]$), is smaller than the maximum value allowed for detection, $\delta_{\rm max}$. This value depends on the NTD material and the charge $z$ of the incident particle. In our case, i.e.\ CR-39 NTDs and $z=1$, $\delta_{\rm max}(\beta\simeq0.15) \simeq 0^{\circ}$, which means that staus travelling faster than $\beta\simeq0.15$ will not be detected.

In Fig.~\ref{fg:inc} we show the distribution of the incidence angle $\delta$ corresponding to the Run-2 geometry. The distribution is obtained through Monte Carlo event generation assuming $m_{\glu} = 1.2~\tev$,  $m_{\glu} - m_{\none} = 30~\gev$ and $m_{\none} - m_{\stau} = 1~\gev$. As can be seen, the stau has an incidence angle smaller than $25^\circ$ about a half of the time, which requires $\beta \lesssim (0.08 \div 0.15)$ to be detected by the NTD.
\begin{figure}[htb]
\centering
\includegraphics[width=0.95\linewidth]{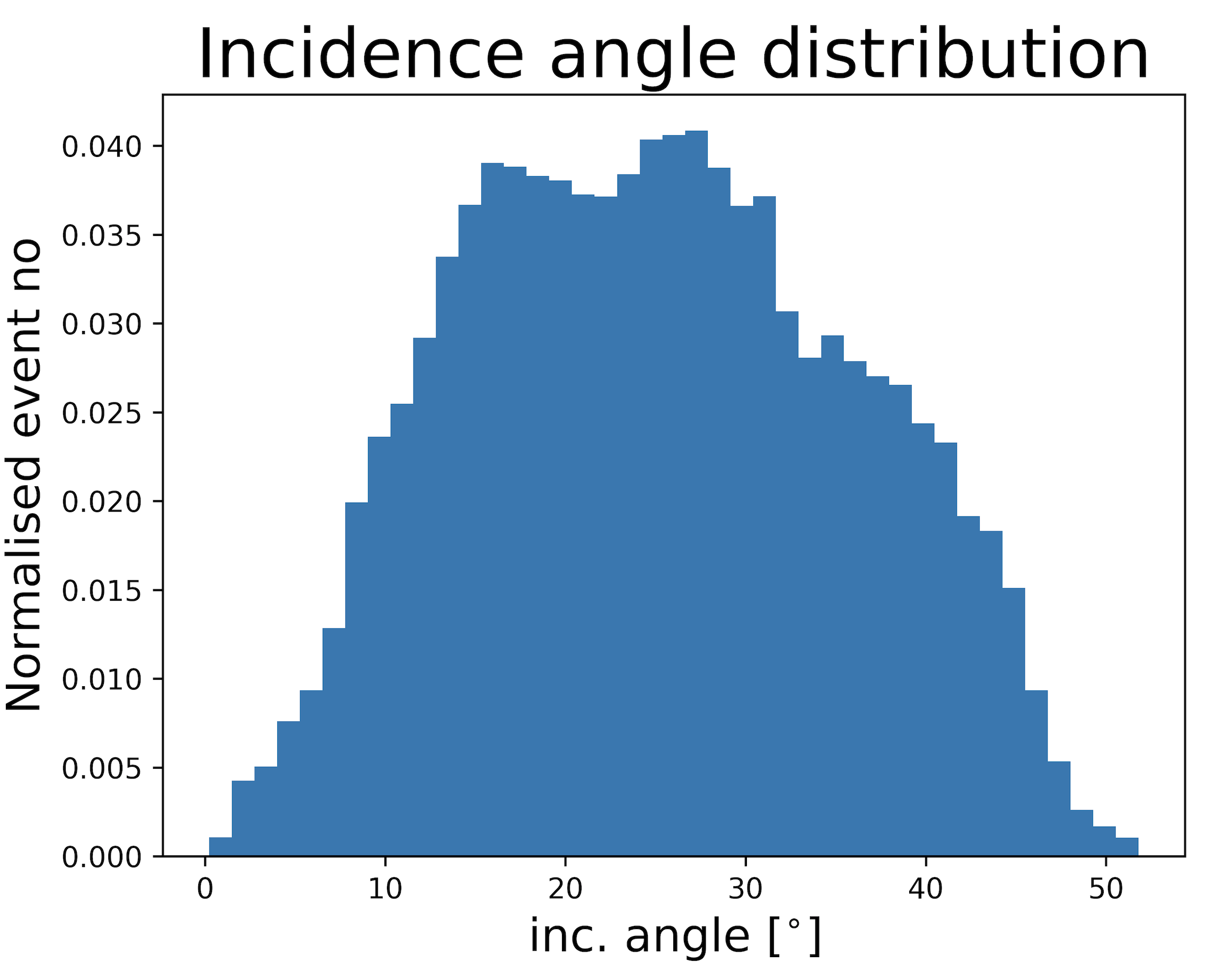}
\caption{\label{fg:inc} The distribution of the incidence angle between the \stauu and an NTD panel assuming the Run-2 NTD geometry.}
\end{figure}
For particles of low $z$, the maximum tilt allowed for the detection of NTD etch-pits is rather low~\cite{Manzoor:2007}, providing strong motivation for an NTD configuration with the minimum possible incidence angle. Therefore, if the NTD panels are installed in the cavern in such a way so that they ``face'' the interaction point, the MoEDAL reach is expected to be improved with respect to the Run-2 geometry. Such a consideration would also have a positive impact on searches for doubly charged Higgs bosons~\cite{Chiang:2012dk} or fermions. Of course, the implementation of this idea relies upon the mechanical implications it will have in the cavern. 

In order to have an estimate for this improved NTD geometry, we also consider in this study an ``ideal'' spherical detector where the incidence angle is $\delta=0$ for every particle coming straight from the interaction point. The realistic detector response for Run-3 is expected to be somewhere between the two extreme cases.

\subsection{Analysis and results}\label{sc:analysis}

We estimate the expected number of signal events by
\begin{linenomath*}
\begin{equation}
N_{\text{sig}} = \sigma_{\glu} \cdot {\cal L} \cdot \epsilon,
\end{equation}
\end{linenomath*}
where $\sigma_{\glu} \equiv \sigma(pp \to \glu \glu)$ is the gluino production cross-section, $\cal{L}$ is the integrated luminosity and $\epsilon$ is the efficiency. From the above consideration, the efficiency can be estimated by the Monte Carlo (MC) simulation as
\begin{linenomath*}
\begin{equation}
\epsilon = \left \langle 
\sum_{i=1,2} P_{\rm NTD}({\bf p}_i)
\cdot \Theta \left( \delta_{\rm max}(\beta_i) - \delta_i  \right)
\right \rangle_{\rm MC},
\end{equation}
\end{linenomath*}
where ${\bf p}_i$, $\beta_i$ and $\delta_i$ are the momentum, velocity and incidence angle of $i$-th neutralino and stau,
$\Theta(x)$ is the step function ($\Theta(x) = 1$ for $x > 0$ and 0 otherwise)
and $\left \langle \cdots \right \rangle_{\rm MC}$ represents the Monte Carlo average.
Due to the extremely low background of the analysis, the observation of even one sole event ($N_{\rm sig} = 1$) would be significant enough to raise interest, while two events ($N_{\rm sig} = 2$) may possibly mean a discovery. Both cases are considered in the analysis.

\begin{figure}[htb]
\centering
\includegraphics[width=\linewidth]{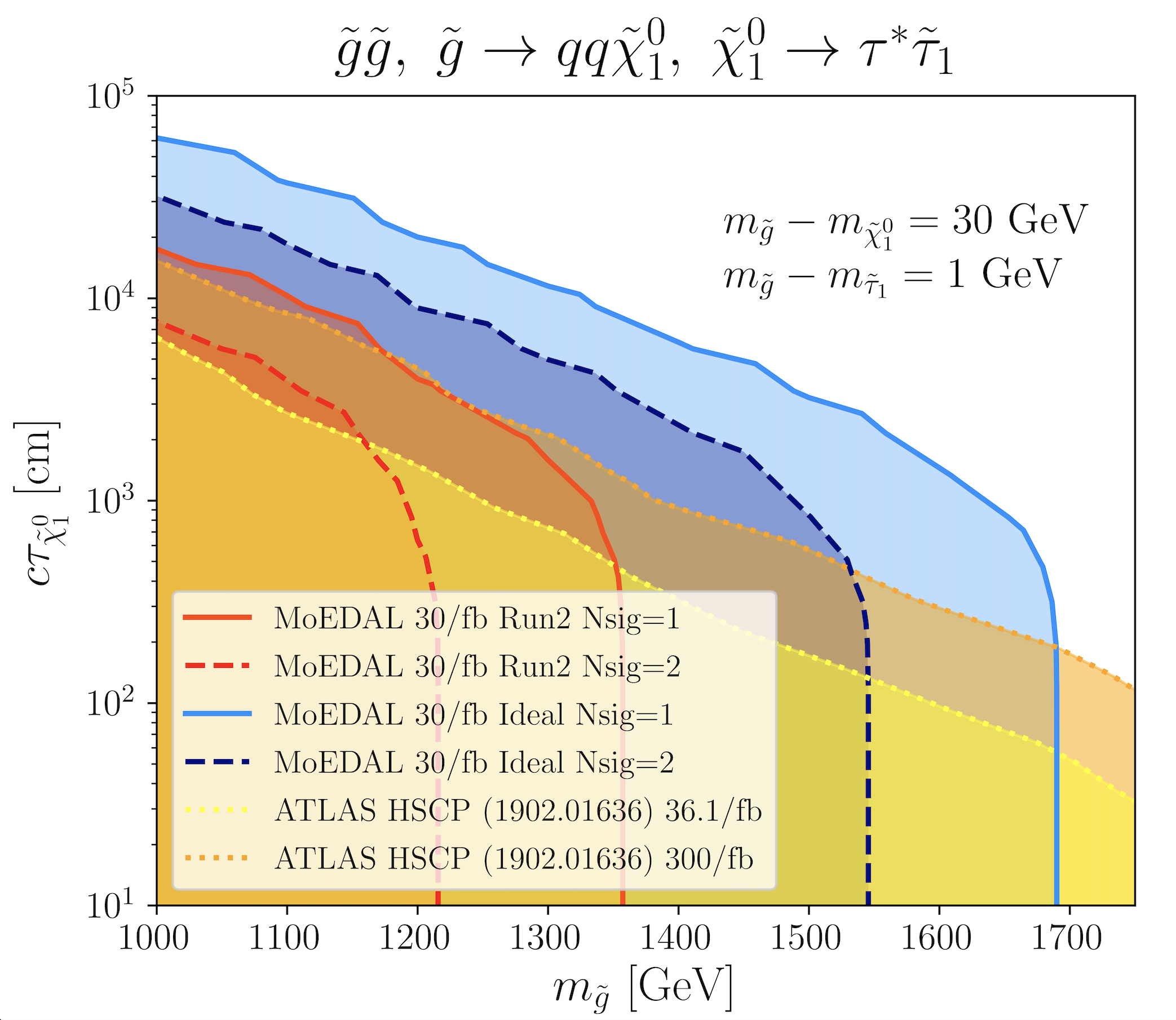}
\caption{\label{fg:limit} 
The sensitivity of MoEDAL, $N_{\rm sig} = 1$ (solid) and $N_{\rm sig} = 2$ (dashed), in the $m_{\glu}$ vs.\ $c \tau_{\none}$ plane for the $pp \to \glu \glu$ production followed by $\glu \to q \bar q \none$, $\none \to \tau^* \stau$. We fix the mass splitting as $m_{\glu} - m_{\none} = 30~\gev$ and $m_{\none} - m_{\stau} = 1~\gev$. Two NTD geometries are considered: the Run-2 discussed in the text (red) and an ideal geometry with all NTD panels facing the interaction point (blue). The region below the dotted yellow contour is excluded by the current ATLAS HSCP analysis with ${\cal L} = 36.1~\ifb$~\cite{Aaboud:2019trc}, while the dotted orange contour represents the projection of this analysis to Run-3 luminosity ${\cal L} = 300~\ifb$. For MoEDAL ${\cal L} = 30~\ifb$ is assumed for Run-3.}
\end{figure}

In Fig.~\ref{fg:limit}, we show the region of $N_{\rm sig} = 1$ (solid lines) and $N_{\rm sig} = 2$ (dashed lines) in the $m_{\tilde g}$ vs.\ $c \tau_{\tilde \chi_1^0}$ plane. We show both geometry scenarios: the (conservative) actual geometry for Run-2 and the ideal spherical one. We assume ${\cal L} = 30~\ifb$, which may be achievable for MoEDAL at the final stage of Run-3, planned to last from 2021 to 2024. 

On the same plot, we superimpose the current limit (dotted yellow) obtained by recasting the ATLAS HSCP analysis~\cite{Aaboud:2019trc} to the simplified model under study. We also show (dotted orange) the projection of this limit to the Run-3 luminosity, ${\cal L} = 300~\ifb$, obtained by simply assuming that the signal and background scale in the same way. We stress here that we do not consider any possible future improvements in the ATLAS (or CMS) analysis, which may enhance its sensitivity either for the di-stau direct production or for more complex topologies, such as the one discussed here. For example, if the pixel hit requirements were significantly relaxed then the ATLAS search would be more powerful than the MoEDAL one across the full parameter space.

As evident, MoEDAL can explore the region of parameter space ($m_{\glu} \lesssim 1.3~\tev$, $c \tau_{\none} \gtrsim 500$~cm), which is currently not excluded. The expected MoEDAL reach is comparable to that of ATLAS HSCP search if the current NTD geometry is used, while the MoEDAL sensitivity may surpass ATLAS's if a nearly  spherical geometry is considered.

The MoEDAL reach clearly shows a different trend than ATLAS (and CMS): MoEDAL may cover larger \none lifetimes, while it is weaker on the \glu mass mostly due the large luminosity needed to overcome the heavier, hence less abundant, gluinos. It is worth stressing here the importance of accessing the same models by both ATLAS and MoEDAL, two experiments with completely different design philosophies, which in case of a positive signal, will help confirm the observation and permit to extract distinct sets of information on the phenomenology.

Finally, we comment on the possible constraint from the prompt gluino search in the jets-plus-missing-transverse-momentum channel. Recently ATLAS and CMS placed stringent lower limits of $1100~\gev$ (ATLAS~\cite{ATLAS:2019vcq}) and $1300~\gev$ (CMS~\cite{Sirunyan:2019xwh}) on the mass of gluino that decays to a stable neutralino ($\glu \to q \bar q \none$) with a compressed mass spectrum $m_{\glu} - m_{\none} \lesssim 50~\gev$. Unlike this case, in our simplified model, the \none is long-lived and decays into a collider-stable \stauu, so this limit cannot be applied directly as it is, since the presence of  displaced and meta-stable staus would affect the trigger efficiency and the estimation of the missing transverse momentum. Although estimating these effects is very complicated and beyond the scope of this paper, it is important to bear in mind that the region with $m_{\glu} \lesssim 1200~\gev$ may be subject to this constraint and already excluded by the prompt-gluino search~\cite{ATLAS:2019vcq,Sirunyan:2019xwh}.

\section{Staus and the anomalous ANITA events}\label{sc:anita}

Before concluding we would like to place our results in the context of some relatively recent discussion on a possible role of SUSY at providing an explanation of the two anomalous events observed by the ANITA (ANtarctic Impulsive Transient Antenna) Collaboration~\cite{Gorham:2016zah,Gorham:2018ydl}.
Although our analysis in the current paper should be viewed completely independently from the ANITA events, it is worth discussing the allowed range of the long-lived $\stauu$ masses accessible to the MoEDAL experiment, in the context of the ANITA events, as an additional motivation for such searches at colliders.

The ANITA  experiment is a balloon-borne detector designed to study ultra-high-energy (UHE) cosmic neutrinos by detecting the radio pulses emitted by their interactions with the Antarctic ice sheet. ANITA recently reported two anomalous events, which resemble air showers initiated by energetic ($\sim 500~\pev$) particles that emerge from the ice moving upwards with large elevation angles (of order $\sim -30^\circ$ below the horizon). These events lack phase inversion. Moreover, such high energy events appear to be in tension with observations by the IceCube detector~\cite{Aartsen:2017mau,Aartsen:2020vir,Safa:2019ege}, which adds to the mystery. Ordinary neutrino-interaction explanations for these anomalous events are excluded~\cite{Romero-Wolf:2018zxt}. More mundane explanations associated with the structure of the Antarctic subsurface have been proposed ~\cite{Shoemaker:2019xlt}.

On the other hand, explanations involving BSM physics have also been 
proposed~\cite{Cline:2019snp,Anchordoqui:2019utb}, including heavy dark matter models~\cite{Hooper:2019ytr} and SUSY~\cite{Dudas:2018npp,Fox:2018syq,Connolly:2018ewv,Collins:2018jpg}. Supersymmetry constitutes, in our opinion, one of the best proposed explanations of these events to date. 

Several of these SUSY explanations involve the production of a long-lived right-handed \stauu (\staur) NLSP~\cite{Albuquerque:2003mi,Ando:2007ds,Fox:2018syq,Connolly:2018ewv}, which in most cases decays to a $\tau$ lepton and a gravitino, if a GMSB model is assumed~\cite{Albuquerque:2003mi,Ando:2007ds,Fox:2018syq}. The \staur can be produced in interactions of nucleons with ultra-high-energy cosmic neutrinos of energies $\sim 1~\eev$. Then, under certain conditions, namely small (less than 100~pb) interaction cross sections of the \staur with the nucleons,  relatively low ionisation and appropriate energies and lifetimes, the resulting \staur can propagate undisturbed for almost the entirety of the Earth's interior until it decays to a $\tau$ lepton and \gra just before it emerges from Earth's surface:
\begin{linenomath*}
\begin{align}\label{staudecay}
\staur \,  \rightarrow \,  \tau \, \gra.
\end{align}
\end{linenomath*}
 The proper lifetime of the \staur that ensures its undisturbed propagation through the Earth's interior from the production point, roughly a distance of order of the Earth's radius $\sim 6000$~km, at energies $\sim 1~\eev$, which, for $m_{\stauu} \simeq 1~\tev$, corresponds to a Lorentz factor $\gamma \sim 10^6$, can thus be estimated to be
 \begin{linenomath*}
 \begin{align}\label{proplife}
 c\tau \, \gtrsim \, 6 \times 10^6 \, \gamma^{-1}~{\rm m} \, \simeq \, 6~{\rm m},
 \end{align}
 \end{linenomath*}
which is long-lived enough for the \stauu to reach and produce high ionisation in the MoEDAL detector. On the other hand, from theoretical models one can estimate that the \staur proper lifetime for mass $m_{\staur}$ is of order~\cite{Fox:2018syq,Albuquerque:2003mi, Ando:2007ds} 
 \begin{linenomath*}
 \begin{align}\label{lifetime} 
 \tau \simeq 10(m_{\staur}/ 500~\gev)~{\rm ns}. 
\end{align}
\end{linenomath*}
Thus, we observe from \eqref{proplife} and \eqref{lifetime}, that, in such scenarios, the ANITA shower-like events are initiated by the hadronic decays of the $\tau$ leptons, and can be produced by \stauu's of mass 
\begin{linenomath*}
\begin{align}\label{staumass}
500~\gev \,  \lesssim m_{\staur} \lesssim \, 1~\tev. 
\end{align}
\end{linenomath*}
This mass range of these \stauu's are in the relevant advantageous range for MoEDAL, as much as for ATLAS and CMS, SUSY searches. The above features are actually generic for any BSM particle with the above properties, not only a $\staur$. 

However, such dominant production mechanisms for the ANITA air showers through hadronic decays of $\tau$ leptons leads to the generic prediction of having similar events in IceCube~\cite{Collins:2018jpg, Huang:2018als}, which have not been detected as yet. This issue could be resolved in RPV models~\cite{Collins:2018jpg}, where sleptons or squarks with mass of order of a \tev produced during the interactions of \eev cosmic neutrinos with nucleons decay ({\it cf.}~\eqref{staudecay}) into a light long-lived bino $\none$ with mass of $\mathcal O(1~\gev)$ and RPV couplings of $\mathcal O(0.1)$. The latter survives propagation through the Earth, before decaying into neutrinos, charged leptons and/or quarks, thus producing upgoing air showers in the neighbourhood of the ANITA balloon. Such models escape the IceCube non-observation mystery by the fact that only a fraction of events proceeds via $\tau$ lepton decays, which would lead to ice-penetrating charged leptons. See also Ref.~\cite{Fox:2018syq} for related discussions.

SUSY models involving long-lived \stauu's have been the focus of our previous discussion. This implies, that the ANITA events could be confirmed/discarded by MoEDAL or ATLAS/CMS searches for long-lived charged particles. In fact, the scenario described in the previous section would be suitable to explain the ANITA events. In our case, we have a long-lived \stauu produced strongly at the LHC though the chain~\eqref{eq:cascade}. The main difference with the standard \stauu explanation of ANITA events is the presence of a long-lived \none degenerate in mass with the \stauu. This fact modifies some features of the event, most importantly, the prediction for the elevation angle of ANITA events.

As an example we take a typical event observable at MoEDAL but not with the ATLAS analysis: $m_{\glu} = 1.3~\tev$,  $m_{\glu}-m_{\none} = 30~\gev$, $m_{\none} - m_{\tilde \tau} = 1~\gev$ and $c \tau_{\none} = 5$~m. Long-lived staus are produced in the Earth's crust by interaction of the ultra-high energy cosmic neutrino with an Earth nucleon at rest. The dominant production chain would involve chargino exchange in the $t$-channel: 
\begin{linenomath*}
\begin{align}\label{nuq}
 \nu_\tau q \to \tilde \tau_{L} \tilde q &\to \left ( \none \tau \right) \left ( \none q \right) \to \left  ( \staur \tau^* \tau \right) \left ( \staur \tau^* q\right). 
\end{align}
\end{linenomath*}

The calculation of the emergence angle is completely analogous to the calculation in Refs.~\cite{Romero-Wolf:2018zxt,Anchordoqui:2019utb} with the addition of an intermediate long-lived \none. In this case, the \none energy degradation is much lower and can be neglected before the \none decays to \staur. After this decay, the calculations in~\cite{Romero-Wolf:2018zxt,Anchordoqui:2019utb} applies and the emergence angle is obtained simply adding $l_\chi = \gamma c \tau_\chi$ to the path distance in Earth calculated for the prompt \stauu, hence, tilting the angle to slightly larger values. However, this change in the emergence angle could always be adjusted with a shorter \stauu lifetime. 

\section{Conclusions and outlook}\label{sc:concl}

We performed a feasibility study on the detection of massive metastable supersymmetric partners with the MoEDAL experiment in a complementary way to ATLAS. Direct production of heavy (hence slow-moving) fermions with large cross section (thus via strong interactions) is the most favourable scenario for MoEDAL. 

MoEDAL is mostly sensitive to slow-moving particles ($\beta \lesssim 0.2$) unlike ATLAS/CMS suitability for faster ones, yet the less integrated luminosity it receives at IP8 remains a limiting factor for simple scenarios. Nonetheless, the results presented here appear to be promising for more complex topologies, e.g.\ those with a neutral LLP in the decay chain. MoEDAL can cover part of the parameter space in such, to a certain extent, elaborate scenarios, which are currently unconstrained by ATLAS and CMS, yet may be probed in the future if some selection criteria are omitted from their respective analyses. 

Even for SUSY models observable by both ATLAS/CMS and MoEDAL, the added value of MoEDAL would remain, since it provides a coverage with a completely different  detector and analysis technique, thus with uncorrelated systematic uncertainties. Indeed, should an excess of events be observed by ATLAS or CMS, good determination of the new particle velocity and mass would be possible under the assumption of unit electric charge. On the other hand, the etch-cone shape of a particle detected in MoEDAL NTDs can provide information on its charge and energy~\cite{Patrizii:2010jla}. The velocity can only be constrained by a maximum value depending on the charge and the (measurable) incidence angle.

We also make a potential connection between the MoEDAL-friendly range of the parameter space of the SUSY models discussed here with that required for an explanation of the ANITA anomalous events, with the caveat though that the latter may admit more mundane explanations, and also the fact that IceCube has not observed similar events. 

More effort is needed towards the exploration of realistic SUSY scenarios where the studied simplified topologies occur naturally. So far, we have only considered sleptons as the metastable particles that interact directly with the MoEDAL detectors; R-hadrons, and possibly charginos, are other possibilities worth examining in the future.  

\begin{acknowledgements}

We would like to thank colleagues from the MoEDAL Collaboration for discussions and comments and in particular J.~Bernabeu, J.~R.~Ellis and A.~de~Roeck, L.~Patrizii and Z.~Sahnoun. The work of JM, VAM, NEM, RRA, AS and OV is supported in part by the Generalitat Valenciana (GV) via a special grant for MoEDAL and by Spanish and European funds under the projects FPA2017-85985-P, FPA2017-84543-P and PGC2018-094856-B-I00 (MCIU/AEI/FEDER, EU). VAM, AS and OV acknowledge support by the GV via the Project PROMETEO-II/2017/033. The work of JM is supported by the GV through the contract APOSTD/2019/165. VAM acknowledges support by a 2017 Leonardo Grant for Researchers and Cultural Creators, BBVA Foundation. The work of NEM is supported in part by the UK Science and Technology Facilities research Council (STFC) under the research grants ST/P000258/1 and ST/T000759/1 and by the COST Association Action CA18108 ``{\it Quantum Gravity Phenomenology in the Multimessenger Approach (QG-MM)}''. NEM also acknowledges a scientific associateship as \emph{Doctor Vinculado} at IFIC-CSIC-Valencia University, Spain. JLP acknowledges support by the Natural Science and Engineering Research Council of Canada via a project grant, by the V-P Research of the University of Alberta and by the Provost of the University of Alberta. The work of KS is partially supported by the National Science Centre, Poland, under research grant 2017/26/E/ST2/00135 and the Beethoven grant DEC-2016/23/G/ST2/04301.  

\end{acknowledgements}
\bibliographystyle{JHEP}   
\bibliography{SUSY_MoEDAL_2019}

\end{document}